\documentclass{article}
\usepackage[affil-it]{authblk}
\usepackage{amssymb}
\usepackage{subfig}
\usepackage{amsmath, amssymb}
\usepackage{hyperref}
\usepackage{url}
\usepackage{graphicx}
\usepackage{mathtools}
\usepackage{tikz}
\usepackage{pgfplots,siunitx}
\usepackage{booktabs}
\usepackage{xcolor}
\usepackage[T1]{fontenc}
\usepackage{siunitx}
\usepackage{tgheros,textcomp}
\usepackage{pgfplotstable}
\pgfplotsset{compat=newest}
\usepgfplotslibrary{units}

\usepackage{colortbl}
\usepackage{amsmath}
\usepackage{xcolor}
\usepackage{graphicx}
\usepackage{longtable}

\newcommand*\polysite{\url{http://polymorph.units.it}}
\newcommand*\defeq{\stackrel{\mathclap{\tiny\normalfont\mbox{def}}}{=}}
\newcommand*\maxH{\hat{H}}

\colorlet{tableheadcolor}{gray!25} 
\newcommand{\headcol}{\rowcolor{tableheadcolor}} %
\colorlet{tablerowcolor}{gray!10} 
\newcommand{\rowcol}{\rowcolor{tablerowcolor}} %
\colorlet{tablerescolor}{gray!20} 
\newcommand{\rescol}{\rowcolor{tablerescolor}} %

\newcommand{\gen}[3][A]{{#1}^{\textrm{#3}}({#2})}       
\newcommand{\Ain}[1][\mathcal{L}]{\gen[I]{#1}{}}
\newcommand{\Aout}[1][\mathcal{L}]{\gen[O]{#1}{}}

\newcommand*{\rulefiller}{%
  \arrayrulecolor{tableheadcolor}
  \specialrule{\heavyrulewidth}{0pt}{-\heavyrulewidth}
  \arrayrulecolor{black}
}

\colorlet{nKBColor}{gray}
\colorlet{KBColor}{darkgray}

\colorlet{P3SpellerColor}{lightgray}
\colorlet{P1MColor}{gray}
\colorlet{P2MColor}{darkgray}
\colorlet{P3SpellerBorderColor}{lightgray}
\colorlet{P1MBorderColor}{gray}
\colorlet{P2MBorderColor}{darkgray}

\colorlet{nKBColor}{blue}
\colorlet{KBColor}{red}

\colorlet{P3SpellerColor}{green}
\colorlet{P1MColor}{blue}
\colorlet{P2MColor}{red}
\colorlet{P3SpellerBorderColor}{green}
\colorlet{P1MBorderColor}{blue}
\colorlet{P2MBorderColor}{red}

\newcommand{\subject}[1]{\textbf{S#1}}
\newcommand{\lang}[1]{\textbf{#1}}

\newcommand{\English}{\lang{En}}
\newcommand{\German}{\lang{Ge}}
\newcommand{\French}{\lang{Fr}}
\newcommand{\Italian}{\lang{It}}
\newcommand{\Finnish}{\lang{Fi}}
\newcommand{\Hungarian}{\lang{Hu}}
\newcommand{\ItalianC}{\lang{It*}}

\newcommand{\Csymb}{\Sigma_{c}}
\newcommand{\Psymb}{\Sigma_{p}}
\newcommand{\Msymb}{\Sigma_{m}}

\newcommand{\nrows}{h} 
\newcommand{\ncols}{w} 

\usepackage{ifthen}

\newcommand{\iratesymb}{r}
\newcommand{\aratesymb}{R}
\newcommand{\ARsymb}{D}

\newcommand{\ItNotation}[2]{\ifthenelse{\equal{#2}{}}{#1}{#1_{#2}}}

\newcommand{\irate}[1][]{\ItNotation{\iratesymb}{#1}} 
\newcommand{\arate}[1][]{\ItNotation{\aratesymb}{#1}} 
\newcommand{\AR}[1][]{\ItNotation{\ARsymb}{#1}} 
\newcommand{\oAR}[1][]{\ItNotation{\overline{\ARsymb}}{#1}}

\graphicspath{{./figures/}}

\title{PolyMorph: Increasing P300 Spelling Efficiency by Selection Matrix Polymorphism and Sentence-Based Predictions}

\author[1]{Alberto Casagrande%
  \thanks{Electronic address: \texttt{acasagrande@units.it}; Corresponding author}}
\author[2]{Joanna Jarmolowska%
  \thanks{Electronic address: \texttt{fabasia@libero.it}}}
\author[2]{Marcello Turconi%
  \thanks{Electronic address: \texttt{marcello.turconi@libero.it}}}
\author[2]{Pierpaolo Busan%
  \thanks{Electronic address: \texttt{pbusan@units.it}}}
\author[1]{Francesco Fabris%
  \thanks{Electronic address: \texttt{ffabris@units.it}}}
\author[2]{Piero Paolo Battaglini%
  \thanks{Electronic address: \texttt{battagli@units.it}}}

\affil[1]{Dept. of Mathematics and Geosciences, University of Trieste, Via Valerio 12/1, 34127 Trieste, Italy}
\affil[2]{Dept. of Life Sciences, University of Trieste,  Via Weiss 2, 34128 Trieste, Italy}

\date{}

\begin{document}

\maketitle

\begin{abstract}
P300 is an electric signal emitted by brain about 300 milliseconds
after a rare, but relevant-for-the-user event. One of the applications of this signal
is sentence spelling that enables subjects who lost the control of their motor
pathways to communicate by selecting characters in a matrix containing all the 
alphabet symbols. Although this technology has made considerable progress 
in the last years, it still suffers from both low communication rate and high error 
rate. This article presents a P300 speller, named PolyMorph, that introduces two major 
novelties in the field: the selection matrix polymorphism, that reduces the 
size of the selection matrix itself by removing useless symbols, and 
sentence-based predictions, that exploit all the spelt characters of a sentence 
to determine the probability of a word. 
In order to measure the effectiveness of the presented speller, we describe two sets of tests:  
the first one \emph{in vivo} and the second one \emph{in silico}.  The results of these experiments 
suggest that the use of PolyMorph in place of the na\"ive character-by-character speller both
increases the number of spelt characters per time unit and  
reduces the error rate. 
\end{abstract}



\section{Introduction}

Brain computer interface (BCI) technology allows individuals with motor disabilities to establish 
a new channel of non-muscular communication with the surrounding environment. 
Virtual keyboards ruled by computers that collect and interpret user brain signals have been investigated since the rise of this technology. 
The cerebral activity piloting such keyboards can be revealed with non-invasive methods such as electroencephalography (EEG)~\cite{1454811,Wolpaw:1991uq}. 
The most frequently studied applications for BCI use the P300 component of the event-related potential (ERP) 
that manifests following a subject response to an external stimulus~\cite{Sutton:1965xy,Farwell:1988kx,Duncan:2009tx}. 
The P300 wave has a positive potential ($>$10 $\mu V$) that appears only after the presentation 
of an expected or rare stimulus, and has a characteristic distribution in posterior EEG signals 
(centro-parieto-occipital)~\cite{Squires:1975uq}. 
The row-column (RC) speller proposed by Farwell and Donchin in 1988 uses the P300 wave and allows 
for sequential selection of a character within a matrix of rows and columns~\cite{Farwell:1988kx}. Guger \emph{et al.} reported that the 89\% of subjects were able to obtain 
an accuracy of 80-100\% with the P3Speller~\cite{Guger:2009ys}. However, the RC method has several limitations, 
one of which concerns the interface itself, which is very tiring for the user, causing a rapid decline in performance. 
Moreover, selection of a single target can require a relatively long time: 3-8 selections per minute can be obtained with a P300-based BCI~\cite{Ryan:2011nx}. 
Thus, compared to the 150 words per minute that are produced using normal speech~\cite{Maclay59}, 
communication through a BCI is much slower and requires significant user attention even to communicate a simple message.
Considering the need to increase the efficacy of communication using a BCI, 
several groups have attempted to improve its performance with P300-based applications 
that reduce the number of errors while increasing its velocity. Some of these studies have focused on the P300 speller 
paradigm~\cite{Sellers2006242,Townsend:2010it,Kaufmann:2011hc}, 
with the aim of enhancing the classification of the signal~\cite{Krusienski:2008uq} to increase the velocity of communication. 
Salvaris and Sepulveda investigated the effects of visual modification of the P300 speller BCI paradigm by introducing 
differences in background color, size and style of symbols, and size of background on the display~\cite{Salvaris:2009dn}. 
In that study, although no single visual protocol was best for all subjects, performance could nonetheless be improved 
using a white background visual protocol; the worst performance was seen with small symbol size. 
The group of Allison studied the effects of different matrix sizes on P300 amplitude, accuracy, and 
performance~\cite{Allison:2003qe}. 
The results indicated that larger matrices evoked a larger P300 amplitude than a smaller matrix. 
Among studies that have focused primarily on reducing errors, the adjacency problem, stating that some involuntary selections 
depend on the distance between characters, has been investigated~\cite{fazel2011}. 
In this context, 
a paradigm based on regions was proposed~\cite{Fazel-Rezai:2008fe}. 
This paradigm, thanks to its graphic interface, minimized the 
effects of overcrowding of stimuli and the adjacency problem.
More recently, a predictive spelling system has been introduced. Ryan \emph{et al.}, for example, studied a predictive spelling 
program (PS) 
in which a classic RC paradigm was integrated with suggestions based on prefixes of a particular word~\cite{Ryan:2011nx}. 
An $8 \times 9$ matrix was used, and suggestions were not presented within the selection matrix but in a separate window. 
This system overcame the non-predictive system 
in terms of both average time needed to complete a sentence 
 (12 min 43 sec vs 20 min 20 sec, respectively) and average output characters per minute  (OCM)
 ($\mu = 5.28$, $\sigma = 1.67$ vs $\mu = 3.76$, $\sigma = 0.75$)~\cite{Ryan:2011nx}.
However, the average accuracy of the predictive system was lower than the non-predictive system 
(84.88\% vs 89.80\%, respectively), 
while there were no significant differences in either bit rates (19.39 vs 17.71, respectively) 
or selections per minute (3.71 vs 3.76, respectively). 
Kaufmann \emph{et al.} proposed a different approach which preserved the level of accuracy achieved by the non-predictive speller~\cite{Kaufmann:2012vf}. 
In this case, predicted words and alphabetic characters were presented in the selection matrix at the same time. 
As result, the bit rate (in terms of selections per minute) 
was high for both the predictive and non-predictive systems ($15.7$ vs $15.1$, respectively). Yet, 
the predictive system exhibited a higher true bit rate (in terms of characters per minute) with respect to 
the non-predictive speller ($20.6$ vs $12$, respectively), it required less time to write an entire sentence, and it enhanced the 
OCM ($\mu=3.83$, $\sigma=0.88$ for the predictive speller vs $\mu=2.12$, $\sigma=0.52$ for the non-predictive one\footnote{The average OCM's are not reported in~\cite{Kaufmann:2012vf}. We computed them as ratio between 
the number of characters in the target sentence (i.e., $45$) and the overall time needed to spell the sentence (Figure 3 in~\cite{Kaufmann:2012vf}).})~\cite{Kaufmann:2012vf}.

We developed a P300 speller, named \emph{PolyMorph}, with two main aims: increasing the output characters per minute with respect to the speller proposed in the literature and enhancing the spelling accuracy. 
These goals have been pursued by using classical information theory tools such as \emph{information rate}, \emph{absolute redundancy}, and \emph{adaptive compression}.  PolyMorph distinguishes user language from 
channel code, that is the code used to spell user messages, and it reduces channel code redundancy by transitorily removing symbols that have probability $0$ from the selection matrix. Because of this, 
both the size of selection matrix and the selectable symbols change from selection to selection giving reasons for the speller name. Moreover, PolyMorph adopts a sentence-based prediction system to better model the probability distribution of the 
user language, improve the word forecasting, and enhance the communication rate.  

This work presents Polymorph's features and measures its efficiency by using both \emph{in-vivo} and \emph{in-silico} experiments. 
The collected results are analyzed and compared to those obtained by the state-of-the-art spellers. 

The speller source code, which has been released 
under the GNU GPL license, and all the data obtained during the experiments are available at 
URL \polysite{}.

\section{PolyMorph}

\emph{PolyMorph} is a P300-based speller that adopts the oddball paradigm to identify one symbol in a (possible) square matrix  
of potential targets. Polymorph differs from the other spellers on the approach: while most of them try to reduce the time required to perform 
a single selection, Polymorph exploits information theory to reduce the number of selections due to spell a complete sentence. 

In order to better understand how PolyMorph works, we first need to introduce in Section~\ref{sec:notion} 
some notions and the notation that we use along all the paper. In Section~\ref{sec:poly_desc}, we will detail PolyMorph 
features.

\subsection{Notions and Notation}\label{sec:notion}

In this paper, we adopt the symbol ``\_'' to represent the space character with the aim to reduce ambiguity about the 
presence or absence of it.

A \emph{string} is a sequence, possibly empty, of symbols in the alphabet $\Sigma=[a-zA-Z.?!\_']$. While we can in theory  
support accented characters, we decided at the first instance to map them into pairs of characters in $\Sigma$ to simplify PolyMorph development. For instance, ``\emph{\`E}'', ``\emph{\'o}'', ``\emph{\"a}'', and ``\emph{\ss}''  
are mapped into ``\emph{E'}'', ``\emph{o'}'', ``\emph{a}'', and ``\emph{ss}'', respectively.

If $s_1$ and $s_2$ are two strings, then $s_1+s_2$ is the string obtained by concatenating $s_2$ to $s_1$. 
The string $s_1$ is \emph{suffix} of $s_2$ if and only if there exists a string $s_3$ such that $s_2=s_3+s_1$; $s_1$ 
is \emph{prefix} of $s_2$ if and only if there exists a string $s_3$ such that $s_2=s_1+s_3$. 

A \emph{sentence} is a string whose last symbol is either ``.'', ``?'', or ``!'' and whose shorter prefixes are not sentences, i.e., ``.'', ``?'', and ``!'' can be present only 
as the last symbol. A \emph{word} is a string in the alphabet $[a-zA-Z']$. Let us notice that the string ``\emph{Fermi's}'' is considered as a single word. 
The \emph{suffix word prefix}, or SWP, of a string $R$ is the longest word that is suffix of $R$. 
Analogously, the \emph{suffix sentence prefix}, or SSP, of $R$ is the longest sentence that is suffix of $R$. 
A string $R$ \emph{completes} a SSP $S$ or a SWP $W$ when there exists a prefix of $R$ that 
is suffix of $S$ or $W$, respectively.

Later on, the SWP and the SSP of what has been spelled after $n$ selections from the begin of the session is denoted by $W_n$ and $S_n$, respectively. Whenever the number of selections are not relevant, we may omit it and write simply $W$ and $S$. Let us notice that $W_n$ is always a suffix of $S_n$. 

\subsection{Features}\label{sec:poly_desc}

When a subject uses a brain speller, he is communicating a message (i.e., a sentence) through a channel, whose input is an EEG and the output is a monitor. 
Since the channel is sequential (i.e., it cannot transfer the entire message in a single step), the message has to be encoded to be transmitted and decoded to be read. 
We distinguish between \emph{user language} and \emph{channel code} or \emph{code}: 
the former is the language of the user and it contains all the sentences that may be spelled. The 
latter is the language of the encoded messages and, in principle, it may account a syntax or an alphabet completely different from that of the user language.  
Most of the spellers proposed in the literature to date encode the message to be spelled character by character. They also 
adopt the message alphabet as code alphabet. 
This code is trivial and, due to the redundancies of natural languages, it may waste time and bit space by  
transferring unnecessary symbols.
For instance, all the English words that begin by ``\emph{xylop}'' also begin with ``\emph{xylophone}''. From this point of view, the suffix ``\emph{hone}'' is useless, as 
it brings no further information with respect to ``\emph{xylop}'', and we can avoid it still preserving the meaning of the message. 

Since the '50s, information theory has addressed the problem of reducing the number of bits required to transmit a message and many codes have been proposed in the 
literature and implemented in real word applications so far~\cite{Huffman:1952:MCM,rissanen76,LZ77,LZ78,gbcode,PPMcode}. 
However, these codes are meant to be handled by computer programs and they completely alter the syntactic structure 
of the messages. On the contrary, we would like to produce a human-oriented code that reduces the number of symbols necessary to spell a sentence in some 
natural language and, at the same time, minimizes the cognitive overhead required to achieve this goal.  

We observed that, in some situations, natural languages have an alphabet that is overabundant with respect to the message 
chunk to be communicated and, since, the size of an encoded message depends 
on the cardinality of the code alphabet, this may be a waste of bit space. 
For instance, none of the English words whose first character is ``\emph{k}'' have as second one either ``\emph{t}'' or ``\emph{z}''. 
Thus, in some sense, once we know that the first character is ``\emph{k}'', the symbols ``\emph{t}'' and ``\emph{z}'' are unnecessary and they 
can be transitorily removed from the code alphabet. 

A further improvement in the channel code can be obtained by considering word probabilities. Huffman showed that it 
is possible to minimize the average size of the overall 
encoded message by representing strings with codes whose length is inversely related to their probability: the higher 
the probability, the shorted the code~\cite{Huffman:1952:MCM}. 
Hence, if we associate some of the selection matrix symbols to word suffices that complete the current 
SSP with high probability, we may save both useless selections and spelling time.  

In order to exploit above considerations and improve the efficiency of the channel code, PolyMorph
maintains a knowledge-base (KB) that stores all the words that can be spelled, a set of sentences, and the frequencies of 
both of them. 
The KB is meant to capture the typical expression and the probability distribution of 
the language of a single subject. It can be trained at the begin of the session simply by providing a file containing a set 
of sentences: these sentences, all the words composing them, and their frequencies are automatically evaluated and stored 
into the KB. During the spelling session, PolyMorph updated the KB every time a sentence is completed and,  
at the end of the spelling session, it saves the KB in a file that can be reloaded at the beginning the successive session. 

The sentences stored in the KB are used to estimate the probability that a word completes the current SSP 
(i.e., $P(T + R | Q+T)$ where $Q+T$ is the SSP and $T+R$ is the word).
For instance, if ``\emph{the\_word\_th}'' is the current SSP, its SWP ``\emph{th}'' is the prefix of the words ``\emph{those}'', 
``\emph{the}'', and ``\emph{that}''. 
However, neither ``\emph{the\_word\_the}'' nor ``\emph{the\_word\_those}'' appear to be part of  
an English sentence and there are many chances that none of them have ever been spelled. 
If this is the case, the KB would suggest that 
``\emph{that}'' has a higher probability 
than ``\emph{those}'' and ``\emph{the}'' to be the next word to be spelled. 
Since this word probability depends on the SSP, we called it \emph{SSP probability} and we distinguish it from the 
\emph{SWP probability} that is  the probability to be the next word to be spelled after a SWP. 
Every time that a sentence is completed, the KB updates both its statistics and the statistics of its words and 
refines the probability model of the channel code. 

At each selection, PolyMorph extracts the SWP $W$ of the current SSP $S$ and it builds 
the selection matrix that contains three kinds of symbols: 1) \emph{character symbols} ($\Sigma_c$), 2) \emph{mandatory symbols} ($\Sigma_m$), 
and 3) \emph{prediction symbols} ($\Sigma_p$). 
Each of the character symbols represents a single character $c$ in the user alphabet. Their number is variable since 
PolyMorph includes in the selection matrix 
only those symbols $c$ such that $W+c$ is a prefix for some words in the KB. 
Because of this, both the size of the spelling matrix and the symbols $\Sigma\defeq\Csymb \cup \Msymb \cup \Psymb$ 
contained in it change from selection to selection 
(from which the name PolyMorph). 
Moreover, at the selection of any character symbol $c$, PolyMorph spells the longest string $C$ such that $W+C$ is a 
prefix for all the words that begin by $W+c$ in the KB. 
For instance, if the SWP is ``\emph{xylo}'' and the character symbol ``\emph{p}'' is selected, PolyMorph spells 
``\emph{phone}'' because all the words in the KB that begin with ``\emph{xylop}'' has also ``\emph{xylophone}'' as prefix. 
Since the above selection corresponds to browse a radix tree~\cite{DBLP:journals/jacm/Morrison68} from the node associated to $W$ 
to the one associated to $W+C$ through an edge whose label is $C$, 
we call this feature \emph{label selection}. 

The mandatory symbols are always present in the selection matrix and can be configured as required. For instance, 
they may contain either an undo symbol to 
delete last selection, punctuation, or a pause symbol to temporarily suspend spelling. In the experimental phase that we 
present in this paper, we set as  mandatory symbols 
``\_'' (i.e., the space character), ``.'', ``?'', and an undo symbol that is meant to cancel last selection. 

\begin{figure}[!ht]
\begin{center}
\subfloat[Prediction phase: the most selected/frequent words are predicted and associated 
to unique numeric IDs.]{\includegraphics[width=0.40\textwidth]{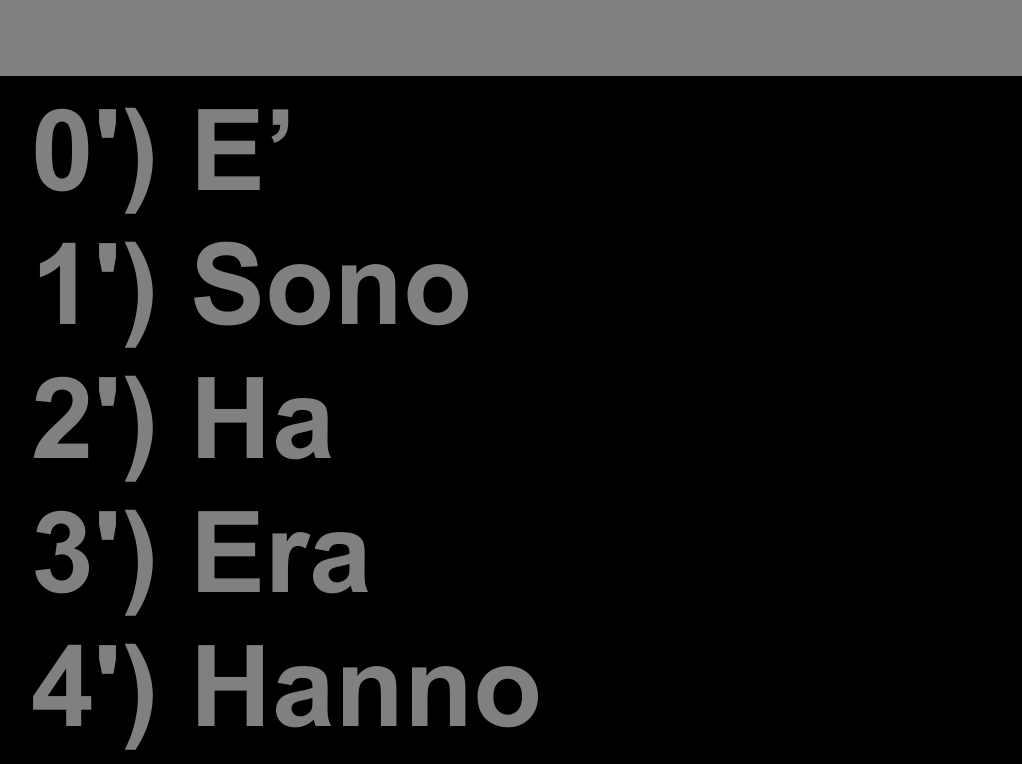}\label{fig:prediction}}
{\hspace{10mm}}
\subfloat[Selection phase: the selection matrix is shown and the P300 measurement proceeds. 
]{\includegraphics[width=0.40\textwidth]{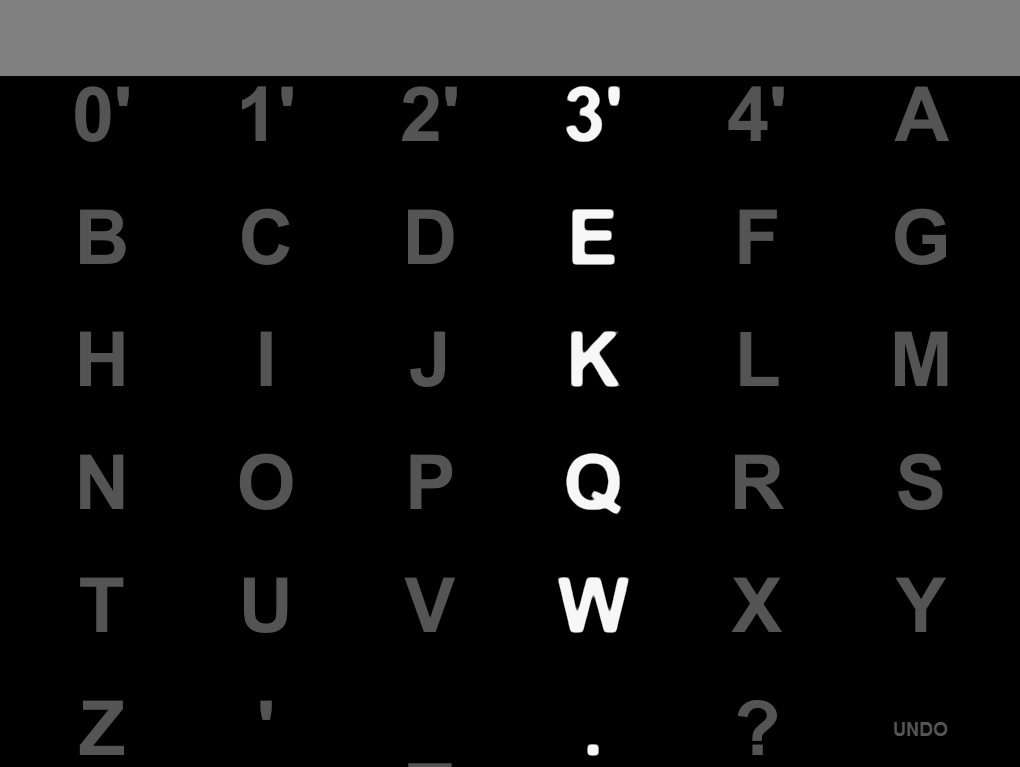}\label{fig:selection}}
\caption{The working cycle of PolyMorph is split into two phases.}\label{fig:polymorph}
\end{center}
\end{figure}

Finally, each of the prediction symbols associates one cell of the selection matrix to a string $R+$``\_''  
such that $W+R$ is a word in the KB that has either 
high SSP probability or high SWP probability. In particular, as long as there are strings for which the 
SSP probability is defined, those strings are presented, then 
high SWP probability strings are considered. The number of prediction symbols depends both on KB and 
on the other symbols in the selection matrix. 
In order to visually present a --potentially-- long string $R+$``\_'' in a single cell in the selection matrix, 
the selection process is split into two phases: the \emph{prediction phase} and the
\emph{identification phase} (see Figure~\ref{fig:polymorph}). The former associate a numeric ID, 
which is the prediction symbol, to the words $W+R$. 
The latter exhibits the selection matrix and performs the
P300 measurements as done in the row-column paradigm. Whenever a suggesting
symbol is selected, the string $R+$``\_'' is spelled. For instance, if ``\emph{the\_word\_th}'' is the current SSP, then 
PolyMorph may associate 
``\emph{that}'' to $0'$ --because of the high SSP probability-- and  ``\emph{the}'' to $1'$ --because of the high SWP probability. 
In such this case, if the user selects $0'$, PolyMorph spells ``\emph{at\_}'' and shows the new SSP ``\emph{the\_word\_that\_}''. 

In order to minimize the ratio between the number of selectable symbols (i.e., $|\Sigma|$) 
and the intensifications required to spell them, both the number of rows ($\nrows$) and columns ($\ncols$) 
of the selection matrix is established dynamically 
and the system ensures that $\nrows \in [\ncols-1,\ncols]$ holds (i.e., the matrix is almost square). 
Because of this, $|\Psymb|$ may change and, as far as there exist enough words in the KB for which 
the current SWP is a prefix, it is the smallest natural number, greater than a user parameter $P_\#$, such that both 
$|\Psymb|+|\Csymb| + |\Msymb|=\nrows*\ncols$ and $\nrows \in [\ncols-1,\ncols]$ hold.

A complete description of algorithms and data structures implemented by PolyMorph is given in~\cite{polymorph}.

\section{Metrics}


In this section, we present some theoretical notions introduced by information theory to evaluate the 
redundancy of a language and its compressibility. Moreover, we both described 
and motivated the metrics used to measure the effectiveness of PolyMorph.

\subsection{Information theoretical notions}\label{sec:inf_theo_notions}
Let $X$ be a \emph{discrete random variable} with possible values in $\{x_1,\ldots,x_n\}$ and 
probability $P_X$. The \emph{entropy} $H(X)$ of $X$ is defined as 
$H(X)\defeq -\sum_{i \in [1,n]}P_X(x_i)*\log{P_X(x_i)}$.

The \emph{joint entropy} is the entropy of a sequence of random variables, e.g., $H(X_1 \ldots X_n)$. 
The \emph{information rate}, or simply rate, $\irate$ is the average entropy per symbol and, in the most general case, it has the 
form $r \defeq \lim_{n \rightarrow \infty}{H(X_1 \ldots X_n)/n}$.

The \emph{maximum entropy} $\maxH(X)$ of a random variable $X$ is the greatest entropy achievable by a random 
variable $Y$ that shares the same support (i.e., the set of values that it can assume) of $X$. It is defined as:
\begin{equation*}
\maxH(X) \defeq \max_{Y\ |\ S(Y)=S(X)} {H(Y)},
\end{equation*}
where $S(X)$ and $S(Y)$ are the supports of $X$ and $Y$, respectively.

The \emph{absolute rate} $\arate$ of a source is the maximum possible rate of information per symbol 
that can be transmitted by it. 
If the alphabet $\Sigma$ of the source does not change during time, it equals $\log{|\Sigma|}$, while, 
in the general case, we defined it as $\lim_{n \leftarrow \infty}{\maxH(X_1 \ldots X_n)/n}$.

Whenever the random variables are clear from the context,  we may write $\irate[n]$ and $\arate[n]$ 
to denote $H(X_1 \ldots X_n)/n$ and $\maxH(X_1 \ldots X_n)/n$, respectively.

\emph{Absolute redundancy} estimates how much redundant is a source and, 
in particular, it evaluates, in average, how many bits per symbol bring no information. More formally, 
the absolute redundancy $D$ of a language is the difference between its 
\emph{absolute rate} and its \emph{information rate}, i.e., $\AR\defeq \arate - \irate$. 

The, so called, \emph{bit rate} estimator~\cite{pierce1980introduction} has been introduced in the literature to evaluate the maximal flow of information per time unit going through the selector. 
The original definition is meant to deal with \emph{stationary} and \emph{memoryless} selectors and has the form:
\begin{equation*}
B\defeq\log_{2}{n}+p\log_{2}{p}+(1-p)\log_{2}{\frac{1-p}{n-1}},
\end{equation*}
where $n$ is the number of selectable objects per selection and $p$ is the probability of correct selection.
%

\subsection{Speller metrics}\label{sec:theo_ana}

%

Bit rate has been successfully used to compare the efficiency of selectors in BCI~\cite{Wolpaw:2000uq}. 
However, 
this metric is strictly related to the 
size of the selection matrix and it does not take into account the efficiency of communication. 
In the particular case of spellers, the transfer rate of a set of messages is always upper bounded by the information rate of their  
language and the average information transmitted by a single selection cannot exceed the information rate 
of the channel code. 
Furthermore, even if the maximal rate of a speller A is greater than that of a speller B, A does not necessarily exploit this advantage 
in transmitting sentences. 
Because of these reasons, 
we consider bit rate suitable neither to evaluate nor to compare speller efficiencies.

On the contrary, \emph{absolute redundancy} (AR) can both relate the absolute rate of a speller and the rate of 
the channel code and measure how many bit are wasted during a single symbol selection. 
The smaller the absolute redundancy, the more efficient the speller is. The effects of AR on the identification phase 
were evaluated by the number of \emph{intensifications per selection and repetition} (ISR) that is equal to $I/(S*R)$ where 
$I$, $S$, $R$ are the total number of 
intensifications, the number of selections necessary to spell the target, and the selection repetitions, respectively.
Above metrics record the redundancy of spelling process, but do not take into account the prediction phase. Thus, 
we also reported both \emph{output characters per minute} (OCM) and the \emph{mean number of selections per minute} (SM) 
to estimate the PolyMorph communication speed-up. 

Finally, the correlation between the use of Polymorph and the number of spelling errors was highlighted by two different metrics: 
the \emph{accuracy} (AC), that 
is number of correct selections 
divided by the total number of selections, and the \emph{errors per character} (EC), that is the ratio between the number of errors and the length of the spelt sentence. 



\section{Methods}\label{sec:methods}

We performed two kinds of tests: an \emph{in-vivo} set of tests, which 
aimed at evaluating the performances of PolyMorph on real users, 
and an \emph{in-silico} set of tests, which unravel the relation  
between the user's language and PolyMorph effectiveness. 
In both the cases, we also tested a classic P3Speller~\cite{Schalk:2004fq} 
to compare the efficiency of the 
two spellers.

\subsection{In-Vivo}

\subsubsection{Partecipants}

The present study considers a total of 10 healthy subjects including 6 males and 4 females. 
Their ages range from 22 to 29 years  with an average of 
24.9 years and a standard deviation of 1.9. All of them were Italian native speakers and 
they were not experts in BCI systems. 
The experimental protocol was prepared in accordance with the Declaration of Helsinki, 
and was approved by the local ethics committee. Moreover, all subjects provided signed informed consent 
forms before the study was initiated.

\subsubsection{Experimental Paradigm}\label{sec:invivo:paradigm}

Each of the considered subjects underwent an experiment consisting of two parts: a \emph{set-up session}, 
which included EEG calibration and a learning phase for the P300 identification procedure, and 
an \emph{on-line session}, which called for some spellings by using both PolyMorph and P3Speller. 

The effectiveness of the proposed system strongly depends on its KB and, because of this reason, 
the on-line session involved both a sentence that was already present 
in it (\emph{target sentence A})
and a one that was not included into the system at beginning of the experiment (\emph{target sentence B}). 

Each sentence was composed twice (\emph{turn 1} and \emph{turn 2}) by using PolyMorph  
to investigate the aftermath of a spelling on its knowledge-base  
and, as a consequence, on the following spellings. 
Since the performances of the standard P3Speller 
depends  neither on a knowledge-base nor on the character distribution of the sentence to be spelled, 
we only required a single spelling of target sentence A for it. 
This results in five different spelling sessions, i.e., two spellings of sentence A by using PolyMorph, 
two spellings of sentence B by using PolyMorph, and one spelling of sentence A by using P3Speller.
With the aim of 
avoiding any bias due to reduced weariness of the subjects, 
the order of these sessions was pseudo-randomized for all participants. 

Since all the subjects involved in our experiments are Italian native speakers, we decided to 
use Italian language during the experiment to avoid any cognitive effort other than 
that required to spell a sentence through P300. 
While the effectiveness of the PolyMorph system depends on the adopted language, 
we expect to obtain the very same conclusions for any language  
that contains some kind of redundancy, as in the case of the natural languages. 

\subsubsection{PolyMorph Knowledge-Base}\label{sec:invivo:knowledge}

In order to initially fill the knowledge-base of PolyMorph, we extracted 
the 200 most commonly used Italian nouns, adjectives, and verbs 
from 
\url{http://telelinea.free.fr/italien/1000_parole.html}. After that,  
we collected the 107.075 most common sentences containing these words from the `\emph{Corpus dell'ita\-lia\-no}' of the \emph{Istituto di Linguistica Computazionale} (\url{http://www.ge.ilc.cnr.it/page.php?ID=moduli_verbi&lingua=it}) and 
we store them into the system. Finally, 
we added 4101 
sentences by randomly selecting them from the web. 
In this way, the initial knowledge-base accounted 111,176 sentences and 51,590 words; in average, the sentences were composed by 37 characters and contained 5.3 words.

In the following, this phrasebook is called \emph{Italian curated phrasebook} (\ItalianC).

\subsubsection{Target sentences}\label{sec:invivo:targets}

The sentences   
``\emph{Piace tanto alla gente.}'' (``People like it very much.") and ``\emph{Sono andato sulla luna.}'' 
(``I went to the Moon.'') 
were chosen as target sentence A and target sentence B, respectively. 
Both these sentences consist of 23 characters, including the spaces and period,  
all their words were contained into the initial knowledge-base (see Sec.~\ref{sec:invivo:knowledge}).

\subsubsection{On-line session}\label{sec:online_session}

The on-line session consisted of spelling the two target sentences (see~Sec. \ref{sec:invivo:targets}).
%
During this session, 
both the duration of the stimulus and the pause between two consecutive stimuli (\emph{inter stimulus interval}) 
were set to 125 ms. 
The time between the appearance of the selection matrix and the first stimulus (\emph{pre-sequence duration}) 
was 3 s in both systems (P3Speller and PolyMorph). The time between the last stimulus and the appearance of the  
selection matrix (\emph{post-sequence duration}) 
was 3 s in the P3Speller, while the duration of the suggested phrase in the PolyMorph system was set to 10 s.

\subsubsection{Data acquisition}
The EEG was registered using a standard cap (Electro-Cap International, Inc.) with 
a modified version of the international placement system to position electrodes. 
The electrodes used for registration were located in the fronto-centro-parietal-occipital cortex, 
and in particular the following electrodes were used: Fz, Cz, P3, Pz, P4, PO7, Oz, and PO8. 
The right mastoid was used as a reference for these electrodes, and the left mastoid was used as the ground. 
Impedance was maintained below 5.0 k\textohm. The signal was amplified and digitized with a Micromed amplifier 
(SAM 32FO fc1; Micromed S.p.A., Italy; analog high-pass filter 0.1 Hz; sampling frequency 256 Hz). 
The signal registered in each channel was processed with a common average reference spatial filter. 
From each EEG channel a data epoch of 800 ms was extracted after presentation of the stimulus. 
BCI2000 software was used to manage the experiment~\cite{Schalk:2004fq} 
(i.e., for presentation of stimuli, collection and elaboration of EEG data, 
and management of the spelling session).

\subsubsection{Set-up session}\label{sec:setup_session}

The set-up session was carried out by using the P3speller software in ``copy mode'' on a $6 \times 6$ matrix. 
In this session, the subjects were requested to spell five alphanumeric strings that were  
composed of four symbols 
each. 
The strings were presented on the upper left side of the computer monitor and 
each successive symbol 
to be selected 
was emphasized by surrounding it by 
parentheses. 
Both the duration 
of each stimulus and the duration of the inter-stimulus interval were 125 ms. Illuminations were organized in sequences 
(in a pseudo-randomized order) of row-column illuminations in which each row and column was illuminated only once. 
A total of 14 sequences were set for each individual symbol. In this way, a total of 28 illuminations 
were obtained for the target stimulus (the number of illuminations for each row plus the number of illuminations for each column) 
and another 140 illuminations for non-target stimuli. Each item was selected for 42 s, 
while the duration of the block lasted about 3 minutes. Considering the pre/post run pause 
and the pre/post stimulus pause, the entire initial session lasted about 15 minutes.

\subsubsection{P300 identification}
The identification of the P300 component was performed during presentation of the target stimulus. The tool ``P300 Classifier''
that is incorporated in the BCI2000 software was used. Stepwise linear discriminant analysis (SWLDA) was used in this 
phase of the experiment. 
This method assumes that the P300 is obtained for one of the six row/column intensifications, 
and 
that the P300 response is not varied compared to the row/column stimuli. 
The resulting classification was taken as the maximum of the sum of the characteristic vectors obtained for the respective rows and columns~\cite{Krusienski:2006os}. 
As a result of the process of discrimination, a transition matrix was generated that estimates the probability of the system 
obtaining a definitive answer (in terms of P300) for each participant. The procedure allows choosing the optimal number of repetitions per stimulus (NRS). 

\subsubsection{On-line parameters}\label{sec:param}

Both the \emph{stimulus duration} (SD) and the \emph{inter-stimuli interval}
(ISI) were set to 125ms. The \emph{pre-sequence duration} (PreS), i.e., 
the time between the appearance of the selection matrix and the first stimulus, was set to 3s for both 
P3Speller and PolyMorph. The \emph{post-sequence duration} (PostS), i.e., the time between the last stimulus and the 
change of selection matrix in P3Speller, was set to 
3s, while the prediction phase lasts 10s (PPD).  
For each subject, we set the minimal NRS that guarantees to maintain the accuracy of the set-up session to 100\%: 
from the first to the tenth subject, NRS was set to 6, 14, 12, 20, 13, 6, 9, 11, 14, and 11, respectively.

\subsection{In-Silico}
The efficiency of PolyMorph depends on the language adopted by users: 
the more redundant it is, the more efficient PolyMorph is expected to be. 
\emph{In-silico} experiments aimed at unravelling this dependency. 

\subsubsection{Experimental paradigm}
We decided to focus on languages based on Latin alphabet and, 
in particular, on English (\English) and German (\German), belonging to the 
Germanic branch of the Indo-European language family, French (\French) and Italian (\Italian), of the Italic branch of the same family, and Finnish (\Finnish) and Hungarian (\Hungarian), two representatives of the  
Uralic languages. The Italian curated phrasebook (\ItalianC) was also considered to better compare \emph{in-vivo} and 
\emph{in-silico} experimental results.

The first analysis estimated how much redundant is the channel code of both P3Speller and PolyMorph varying the 
user language and it 
provided a measure of the spelling efforts wasted due to redundancy. Since the channel code of PolyMorph 
depends exclusively on the KB, we were 
required to build a phrasebook $\mathcal{P}_{\mathcal{L}}$, to store all its sentences in a language-specific KB, 
named ${\textrm{KB}}_{\mathcal{L}}$, and, finally, to 
infer AR, 
for each of the considered user languages $\mathcal{L}$. 

As it concerns the second part of the \emph{in-silico} analysis, we investigated how the efficiency of PolyMorph and 
P3Speller is related to both 
user language and target sentences. For each language $\mathcal{L}$, we built two phrasebooks: $\Ain$, 
whose sentences were all included into 
$\mathcal{P}_{\mathcal{L}}$, and $\Aout$, whose sentences were not included into $\mathcal{P}_{\mathcal{L}}$.
We simulated the behaviour of PolyMorph during the spelling of the sentences in $\Ain$  and $\Aout$ by using $\textrm{KB}_{\mathcal{L}}$ and we recorded 
the sizes of the spelling matrix for all the selections. 

Above experiments were repeated avoiding both the prediction phase and all the prediction symbols with the intent of better understand the relevance of each of the 
PolyMorph features.


\subsubsection{Phrasebooks and knowledge-bases}~\label{sec:insilco:phrasebooks}
We wrote a computer program to collect sentences from the web and we built, for each of  the considered languages $\mathcal{L}$, 
a phrasebook $\mathcal{T}_{\mathcal{L}}$ containing 100,000 --possibly repeated-- sentences. 
Any character having an accent, was replaced by the corresponding ASCII 
character and, if it was the last of a word, we added an apostrophe after it. 
Moreover, all the characters that are not in $[a-zA-Z.?]$ were removed. 

The phrasebooks $\Ain$ and $\Aout$ are subsets of $\mathcal{T}_{\mathcal{L}}$ and each of them contains $10,000$ distinct sentences. 
The sentences were included either in $\Ain$ or in $\Aout$ by a computer program. The same program 
built the phrasebook $\mathcal{P}_{\mathcal{L}}$ as the phrasebook that contains all the sentences in $\mathcal{T}_{\mathcal{L}}$, 
but those in $\Aout$. The knowledge-base $\textrm{KB}_{\mathcal{L}}$ was filled by using all the sentences in 
$\mathcal{P}_{\mathcal{L}}$ and all the words in $\Aout$ that are not contained in $\mathcal{P}_{\mathcal{L}}$. 

\begin{table}[!ht]
\begin{center}
\resizebox{\columnwidth}{!}{%
\begin{tabular}{lllllllll}
\toprule
\headcol & \multicolumn{4}{c}{$\Aout$}& \multicolumn{4}{c}{$\Ain$}\\
 \rulefiller\cmidrule(r){2-5} \cmidrule(r){6-9}
\headcol $\mathcal{L}$&Characters&Words&Sentences&C/W&Characters&Words&Sentences&C/W\\
\midrule
\English&$1,104,177$&$180,161$&$10,000$&$6.13$&$1,166,867$&$189,938$&$10,000$&$6.14$\\
\rowcol\German&$1,015,970$&$137,598$&$10,000$&$7.38$&$1,021,348$&$137,450$&$10,000$&$7.43$\\
\French&$1,183,358$&$185,089$&$10,000$&$6.39$&$1,055,790$&$166,292$&$10,000$&$6.35$\\
\rowcol\Italian&$1,132,476$&$172,385$&$10,000$&$6.57$&$1,011,753$&$154,604$&$10,000$&$6.54$\\
\Finnish&$898,813$&$102,291$&$10,000$&$8.79$&$875,585$&$99,874$&$10,000$&$8.77$\\
\rowcol\Hungarian&$914,398$&$146,779$&$10,000$&$6.23$&$953,824$&$151,024$&$10,000$&$6.32$\\
\ItalianC&$356,238$&$55,005$&$10,000$&$6.48$&$383,541$&$59,631$&$10,000$&$6.43$\\
\bottomrule
\end{tabular}}
\caption{$\Aout$'s and $\Ain$'s statistics. The column C/W reports the average characters per word.}
\label{table:statistics} 
\end{center}
\end{table}

\subsubsection{Absolute redundancy evaluation}

Absolute redundancy computation requires to evaluate the entropy of the channel code. 
Unfortunately, accurate estimation of the entropy for natural language is a complex task 
that requires thousands of experiments (see~\cite{shannon51entropy,Schurmann:1996dz,BFMX03}) and, for sure, it goes far beyond the intent of this work. 
Because of this, we decided to approximate its evaluation both by assuming independence between two 
successive words of the user language and by restricting the words to the ones contained in a dataset. 

The absolute rate of the channel code was lower bounded by 
$\log{|\Sigma|}$ (i.e., $\approx 4.95$) for P3Speller and approximated by $\arate[1000]$ for 
PolyMorph. 
In the same way, the rates of the channel code of both PolyMorph and 
P3Speller were estimated as $\irate[1000]$ and 
their absolute redundancies were appraised as 
$\AR[1000]\defeq \arate[1000]-\irate[1000]$ and $\oAR[1000]\defeq \arate-\irate[1000]$, respectively.

\subsubsection{Simulation parameters and other metrics computation}

The NRS was set to the mean of the user NRS's of the \emph{in-vivo} experiments (i.e., $12$).
As concern the remaining parameters, they were left unchanged with respect to whose reported in Section~\ref{sec:param}.

By taking into account this setting and the size of the selection matrix at each selection, which was obtained as output from the simulations, 
we were able to compute ISR, OCM, and SM. In particular, the number of intensifications required by both PolyMorph and P3Speller to spell one
symbol on a $\nrows \times \ncols$ speller matrix is  $N_{i}=(\nrows + \ncols) * \textrm{NRS}$ and the time spent in such a selection is  $N_{i}*\textrm{SD}+(N_{i}-1)*\textrm{ISI}+\textrm{PPD}+\textrm{PreD}$ for PolyMorph and 
$N_{i}*\textrm{SD}+(N_{i}-1)*\textrm{ISI}+\textrm{PostD}+\textrm{PreD}$ for P3Speller.

Since the \emph{in-silico} experiments were carried out automatically, they did not contain spelling errors and, thus, we did evaluate neither AC nor EC for them.

\section{Data Analysis}

\subsection{Dependent variables}


In order to understand the relevance of the data coming from both \emph{in-vivo} and \emph{in-silico} experiments, 
ISR, OCM, SM, AC, and EC were used as dependent variables in repeated-measures analysis of variance (rmANOVA) 
when data were normally distributed (data normality was verified by using Shapiro-Wilk test). 
First, the factors included in the model were related to sentence writing (sentence A) and speller system 
(three levels: turn 1 and 2 of PolyMorph  
and then P3Speller). Moreover, we performed two-way rmANOVA, in this case factors included in the model were related to sentence writing (two levels: 
sentence A and sentence B) and speller system (two levels: turn 1 and 2 for PolyMorph). 
A $p$-value level of $<0.05$ was considered statistically 
significant.  In cases where a significant interaction and/or when an effect representing a main factor was detected,  post-hoc 
analysis was carried out using a Student's t-test. 

When not normally distributed data were present, we used non-parametric methods. In particular, the Friedman test was used for one-way repeated 
measures analysis of variance by ranks. 
In this case, post-hoc analysis was performed by using Wilcoxon Signed Rank Test. 
Statistical tests are always intended as two-tailed. 

\section{Results}
\label{sec:results}

In the remaining parts of this article, we write $\mu$, $\sigma$, and $\rho$ to denote means, standard deviations, and correlations, respectively. 


\subsection{\emph{In-vivo}}\label{sec:results:invivo}

\pgfplotstableread{
Subject	P1A	P2A	P1B	P2B P3Speller P1M P2M
1	0.0	0.0	0.0	0.0	0.0	0.0	0.0
2	0.11	0.0	0.0	0.0	0.07	0.05	0.0
3	0.0	0.0	0.0	0.0	0.13	0.0	0.0
4	0.0	0.0	0.0	0.0	0.07	0.0	0.0
5	0.0	0.0	0.0	0.14	0.17	0.0	0.08
6	0.17	0.0	0.13	0.14	0.28	0.15	0.08
7	0.11	0.0	0.0	0.0	0.1	0.05	0.0
8	0.0	0.0	0.0	0.0	0.0	0.0	0.0
9	0.0	0.0	0.0	0.0	0.13	0.0	0.0
10	0.0	0.0	0.13	0.0	0.13	0.09	0.0

}\vivoerrorseltable

Concerning sentence A, we observed larger OCM in both turn 1 and 2 of PolyMorph session with respect to that obtained by using P3Speller 
($df = 2$; $\chi^2 = 20$; $p$-value $< 0.00005$; Wilcoxon test: $p$-value $=0.00195$ in both cases).
It was also detected 
relevant enhancement of OCM from turn 1 to turn 2 
(Wilcoxon test: $p$-value $= 0.00195$). 

The rmANOVA showed a significant interaction between writing system and spelled sentence ($F_{(1; 9)} = 6.33$; $p$-value $= 0.033$). 
With regard to PolyMorph, we found a substantially increased OCM 
in turn 1 of sentence A compared with the OCM of the same turn of sentence B  
($t_{(9)} = 6.69$; $p$-value $<0.0009$). 
Finally, the OCM obtained during turn 2 is larger than that of turn 1 for both sentence A and B 
($t_{(9)} = 4.2$; $p$-value $<0.002$ and $t_{(9)} = 9.88$; $p$-value $<0.0009$, respectively). 

\pgfplotstableread{
Subject P1A     P2A     P1B     P2B     P3Speller       P1M     P2M
1       6.48    8.94    4.16    8.94    2.51    5.32    8.94
2       2.85    5.03    2.35    5.03    1.07    2.60    5.03
3       4.11    5.65    2.64    5.65    1.06    3.37    5.65
4       2.76    3.79    1.78    3.79    0.78    2.27    3.79
5       3.87    5.32    2.49    3.80    0.88    3.18    4.56
6       3.76    8.94    3.06    6.39    1.07    3.41    7.66
7       3.94    6.92    3.23    6.92    1.40    3.59    6.92
8       4.37    5.01    2.81    6.02    1.54    3.59    5.51
9       3.66    4.28    2.35    5.03    0.93    3.01    4.66
10      4.37    5.12    2.12    6.02    1.15    3.25    5.57

}\vivoOCMtable

\begin{figure}[!h]
\begin{center}
\resizebox{\columnwidth}{!}{%
\begin{tikzpicture}[scale=0.70]
\begin{axis}
[
width=22cm,
height=6cm,
ymajorgrids,
ybar,
xlabel = Subjects,
nodes near coords, 
every node near coord/.append style={rotate=90, anchor=west},
symbolic x coords={1,2,3,4,5,6,7,8,9,10},
xtick={1,2,3,4,5,6,7,8,9,10},
xticklabels={\subject{1},\subject{2},\subject{3},\subject{4},\subject{5},\subject{6},\subject{7},\subject{8},\subject{9},\subject{10}},
ytick={3,6,9},
ylabel = {Mean OCM},
ymin = 0,
ymax=11.5,
legend columns=-1,
legend style={/tikz/every even column/.append style={column sep=0.3cm},at={(0.5,-0.4)}, anchor=north}
]
\addplot[P3SpellerBorderColor,fill=P3SpellerColor] table[y =P3Speller] from \vivoOCMtable;
\addlegendentry{P3Speller};
\addplot[P1MBorderColor,fill=P1MColor] table[y =P1M] from \vivoOCMtable;
\addlegendentry{PolyMorph - Turn 1};
\addplot[P2MBorderColor,fill=P2MColor] table[y =P2M] from \vivoOCMtable;
\addlegendentry{PolyMorph - Turn 2};
\end{axis}
\end{tikzpicture}}
\caption{Mean output characters per minute \emph{in-vivo}: the values 
reported for PolyMorph turn 1 and 2 are the means of the spelling times 
of sentence A and B in turn 1 and turn 2 respectively.}\label{fig:OCM_vivo}
\end{center}
\end{figure}
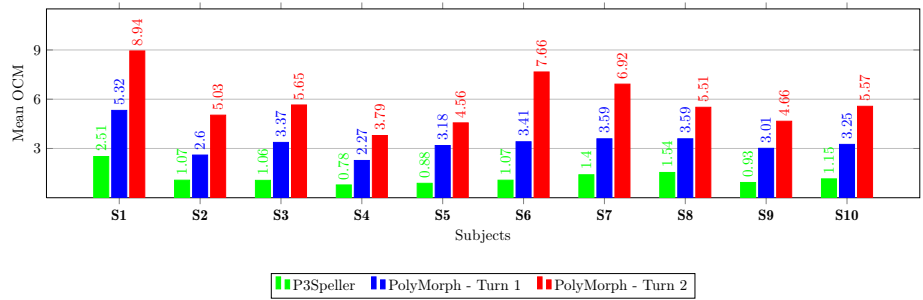


When accuracy was assessed, there were significant differences among conditions 
($df = 2$; $\chi^2 = 6.87$, $p$-value $= 0.032$). Post-hoc analysis show a significant enhancement of accuracy 
when spelling the  sentence A for the $1^\textrm{st}$ time (turn 1) by PolyMorph (and, partially, also for turn 2) 
with respect to that obtained by P3Speller (Wilcoxon test: $p$-value $= 0.039$ and $p$-value $= 0.074$, respectively).

\begin{table}[!h]
\begin{center}
\resizebox{\columnwidth}{!}{%
\subfloat[Accuracy (AC). Data are reported in the format ``(number of correct selections)/(total number of selections)''.\label{table:accuracy}]{
\begin{tabular}{lccccc}
\toprule
\headcol & \multicolumn{3}{c}{Sentence A}& \multicolumn{2}{c}{Sentence B}\\
 \rulefiller\cmidrule(r){2-4} \cmidrule(r){5-6}
\headcol Subj &  PM $1$& PM $2$ & P3S & PM $1$ &  PM $2$ \\
\midrule
\subject{1} & $7/7$ & $5/5$ & $23/23$ & $11/11$ & $5/5$ \\
\rowcol  \subject{2}& $8/9$ & $5/5$ & $25/27$ & $11/11$ & $5/5$\\
\subject{3} & $7/7$ & $5/5$ & $27/31$ & $11/11$ & $5/5$\\
\rowcol \subject{4} & $7/7$ & $5/5$ & $25/27$ & $11/11$ & $5/5$\\
\subject{5} & $7/7$ & $5/5$ & $29/35$&  $11/11$ & $6/7$\\
\rowcol \subject{6} & $10/12$ & $5/5$ & $37/54$ &  $13/15$ & $6/7$\\
\subject{7} & $8/9$ & $5/5$ & $26/29$ &  $11/11$ & $5/5$\\
\rowcol \subject{8} & $7/7$ & $6/6$ & $23/23$ &  $11/11$ & $5/5$\\
\subject{9} & $7/7$ & $6/6$ & $27/31$ &  $11/11$ & $5/5$\\
\rowcol \subject{10} & $7/7$ & $5/5$ & $27/31$ &  $13/15$ & $6/6$\\
\midrule
\rescol Total & $75/79$ & $52/52$ & $272/311$ & 
$114/118$ & $53/55$\\
\bottomrule
\end{tabular}}
\hskip3mm
\subfloat[Errors per character (EC). Data are reported in the format ``(wrong selections)/(number of characters in the sentence)''.\label{table:errors_per_character}]{
\begin{tabular}{lccccc}
\toprule
\headcol & \multicolumn{3}{c}{Sentence A}& \multicolumn{2}{c}{Sentence B}\\
 \rulefiller\cmidrule(r){2-4} \cmidrule(r){5-6}
\headcol Subj &  PM $1$& PM $2$ & P3S & PM $1$ &  PM $2$ \\
\midrule
\subject{1} & $0/23$ & $0/23$ & $0/23$ & $0/23$ & $0/23$ \\
\rowcol \subject{2} & $1/23$ & $0/23$ & $2/23$ & $0/23$ & $0/23$\\
\subject{3} & $0/23$ & $0/23$ & $4/23$& $0/23$ & $0/23$\\
\rowcol \subject{4} & $0/23$ & $0/23$ & $2/23$ & $0/23$ & $0/23$\\
\subject{5} & $0/23$ & $0/23$ & $6/23$&  $0/23$ & $1/23$\\
\rowcol \subject{6} & $2/23$ & $0/23$ & $17/23$ &  $2/23$ & $1/23$\\
\subject{7} & $1/23$ & $0/23$ & $3/23$ &  $0/23$ & $0/23$\\
\rowcol \subject{8} & $0/23$ & $0/23$ & $0/23$ &  $0/23$ & $0/23$\\
\subject{9} & $0/23$ & $0/23$ & $4/23$ &  $0/23$ & $0/23$\\
\rowcol \subject{10} & $0/23$ & $0/23$ & $4/23$ &  $2/23$ & $0/23$\\
\midrule
\rescol Total & $4/230$ & $0/230$ & $41/230$ & 
$4/230$ & $2/230$\\
\bottomrule
\end{tabular}}}
\caption{Accuracy (AC) and errors per character (EC) for PolyMorph turn 1 (PM $1$),  turn 2 (PM $2$),  
and P3Speller (P3S) for \emph{in-vivo} experiments.}
\end{center}
\end{table}


Errors per each character selected were also assessed. Test indicates that the distribution of results significantly differ 
among conditions ($df = 2$;  $\chi^2 = 14.89$, $p$-value $=0.00058$). Post-hoc analysis 
exhibits a significant reduction of errors per each character in both PolyMorph turn 1 and 2 of sentence A 
with respect to errors per each character 
obtained when using P3Speller (Wilcoxon test: $p$-value is  $0.0078$ in both cases). 


\subsection{\emph{In-silico}}\label{sec:results:insilico}

\pgfplotstableread{
Lang	Time_nKB	C_nKB	W_nKB	S_nKB	MTC_nKB	MTW_nKB	MTS_nKB	CPM_nKB	WPM_nKB	SPM_nKB	Time_KB	C_KB	W_KB	S_KB	MTC_KB	MTW_KB	MTS_KB	CPM_KB	WPM_KB	SPM_KB	Time_P3S	C_P3S	W_P3S	S_P3S	MTC_P3S	MTW_P3S	MTS_P3S	CPM_P3S	WPM_P3S	SPM_P3S
English	24937550.125	1104177	180161	10000	22.585	138.418	2493.755	2.66	0.43	0.02	11515342.250	1166867	189938	10000	9.869	60.627	1151.534	6.08	0.99	0.05	46513456.125	1104177	180161	10000	42.125	258.177	4651.346	1.42	0.23	0.01
German	21477311.000	1015970	137598	10000	21.140	156.087	2147.731	2.84	0.38	0.03	8691222.875	1021348	137450	10000	8.510	63.232	869.122	7.05	0.95	0.07	42797736.250	1015970	137598	10000	42.125	311.035	4279.774	1.42	0.19	0.01
French	25375109.500	1183358	185089	10000	21.443	137.097	2537.511	2.80	0.44	0.02	9881060.500	1055790	166292	10000	9.359	59.420	988.106	6.41	1.01	0.06	49848955.750	1183358	185089	10000	42.125	269.324	4984.896	1.42	0.22	0.01
Italian	25516760.000	1132476	172385	10000	22.532	148.022	2551.676	2.66	0.41	0.02	9539286.625	1011753	154604	10000	9.428	61.701	953.929	6.36	0.97	0.06	47705551.500	1132476	172385	10000	42.125	276.738	4770.555	1.42	0.22	0.01
Finnish	19689547.500	898813	102291	10000	21.906	192.486	1968.955	2.74	0.31	0.03	6961666.500	875585	99874	10000	7.951	69.704	696.167	7.55	0.86	0.09	37862497.625	898813	102291	10000	42.125	370.145	3786.250	1.42	0.16	0.02
Hungarian	15148221.625	914398	146779	10000	16.566	103.204	1514.822	3.62	0.58	0.04	9184826.375	953824	151024	10000	9.629	60.817	918.483	6.23	0.99	0.07	38519015.750	914398	146779	10000	42.125	262.429	3851.902	1.42	0.23	0.02
{Italian Curated}	8000037.250	356238	55005	10000	22.457	145.442	800.004	2.67	0.41	0.07	4646009.250	383541	59631	10000	12.113	77.913	464.601	4.95	0.77	0.13	15006525.750	356238	55005	10000	42.125	272.821	1500.653	1.42	0.22	0.04

}\insilicodata

\pgfplotstableread{
Lang	Time_nKB	C_nKB	W_nKB	S_nKB	MTC_nKB	MTW_nKB	MTS_nKB	CPM_nKB	WPM_nKB	SPM_nKB	Time_KB	C_KB	W_KB	S_KB	MTC_KB	MTW_KB	MTS_KB	CPM_KB	WPM_KB	SPM_KB	Time_P3S	C_P3S	W_P3S	S_P3S	MTC_P3S	MTW_P3S	MTS_P3S	CPM_P3S	WPM_P3S	SPM_P3S
English	37921461.250	1104177	180161	10000	34.344	210.487	3792.146	1.75	0.29	0.02	39828620.250	1166867	189938	10000	34.133	209.693	3982.862	1.76	0.29	0.02	46513456.125	1104177	180161	10000	42.125	258.177	4651.346	1.42	0.23	0.01
German	32043448.500	1015970	137598	10000	31.540	232.877	3204.345	1.90	0.26	0.02	31964631.500	1021348	137450	10000	31.297	232.555	3196.463	1.92	0.26	0.02	42797736.250	1015970	137598	10000	42.125	311.035	4279.774	1.42	0.19	0.01
French	38984989.750	1183358	185089	10000	32.944	210.628	3898.499	1.82	0.28	0.02	34868144.375	1055790	166292	10000	33.026	209.680	3486.814	1.82	0.29	0.02	49848955.750	1183358	185089	10000	42.125	269.324	4984.896	1.42	0.22	0.01
Italian	37166896.625	1132476	172385	10000	32.819	215.604	3716.690	1.83	0.28	0.02	33281607.750	1011753	154604	10000	32.895	215.270	3328.161	1.82	0.28	0.02	47705551.500	1132476	172385	10000	42.125	276.738	4770.555	1.42	0.22	0.01
Finnish	25333221.000	875585	99874	10000	28.933	253.652	2533.322	2.01	0.24	0.02	29092310.875	914398	146779	10000	31.816	198.205	2909.231	2.07	0.30	0.02	36884018.125	875585	99874	10000	42.125	369.306	3688.402	1.42	0.16	0.02
Hungarian	30207438.250	953824	151024	10000	31.670	200.017	3020.744	1.97	0.30	0.02	11499202.125	356238	55005	10000	32.280	209.057	1149.920	1.89	0.29	0.05	40179836.000	953824	151024	10000	42.125	266.049	4017.984	1.42	0.23	0.01
{Italian Curated}	11499202.125	356238	55005	10000	32.280	209.057	1149.920	1.86	0.29	0.05	12394496.500	383541	59631	10000	32.316	207.853	1239.450	1.86	0.29	0.05	15006525.750	356238	55005	10000	42.125	272.821	1500.653	1.42	0.22	0.04

}\insilicodatanohints

\pgfplotsset{unit code/.code={\si{#1}}}

PolyMorph reduces the absolute redundancy of the channel code in a significant way with respect to P3Speller for all the considered languages ($\mu\approx0.93$, $\sigma\approx0.11$ smaller than $\mu\approx4.13$, $\sigma\approx0.13$, Wilcoxon test: $p$-value $=0.015626$), but  
the amplitude of this reduction does not depend in a substantial way on the adoption of a specific language. 
The AR reduction affects ISR which is also decreased. However, due to the prediction characters introduced in PolyMorph, this contraction becomes evident exclusively 
when the sentence to be spelt is not present in the KB (see Tables~\ref{table:ISR}). As the matter of facts, 
Wilcoxon test revealed a significant reduction in ISR when using PolyMorph in place of P3Speller in such a case ($\mu\approx11.22$, $\sigma\approx0.37$ smaller than $12.00$,  Wilcoxon test: $p$-value $=0.015626$). 
On contrary, the same test does not highlight differences between PolyMorph and P3Speller in ISR when the spelt sentence is contained in the KB ($\mu\approx12.42$, $\sigma\approx0.57$ vs $12.00$,  Wilcoxon test: $p$-value $=0.09375$), 
but it tops again P3Speller if we take into account the additional row due to prediction ($13.00$ for P3Speller with an additional row, Wilcoxon test: $p$-value $=0.015626$). 

\begin{table}[!h]
\begin{center}
\resizebox{\columnwidth}{!}{%
\begin{tabular}{lccccccccccc}
\toprule
\headcol & \multicolumn{3}{c}{P3Speller}& &\multicolumn{3}{c}{PolyMorph - Pred.} & &\multicolumn{3}{c}{PolyMorph - No Pred.}\\
 \rulefiller\cmidrule(r){2-4} \cmidrule(r){6-8}\cmidrule(r){10-12}
\headcol $\mathcal{L}$&$\arate>$&$\irate[1000]$&$\oAR[1000] $&&$\arate[1000]$&$\irate[1000]$&$\AR[1000]$&&$\arate[1000]$&$\irate[1000]$&$\AR[1000]$\\
\midrule
\English&$4.95$&$0.97$&$3.98$&&$4.85$&$4.01$&$0.84$&&$4.85$&$4.30$&$0.29$\\
\rowcol \German&$4.95$&$0.80$&$4.15$&&$4.82$&$3.91$&$0.91$&&$4.82$&$4.23$&$0.32$\\
\French&$4.95$&$0.85$&$4.10$&&$4.82$&$3.95$&$0.87$&&$4.82$&$4.26$&$0.31$\\
\rowcol \Italian&$4.95$&$0.81$&$4.14$&&$4.81$&$3.99$&$0.83$&&$4.81$&$4.26$&$0.28$\\
\Finnish&$4.95$&$0.73$&$4.22$&&$4.79$&$3.82$&$0.96$&&$4.79$&$4.21$&$0.39$\\
\rowcol \Hungarian&$4.95$&$0.63$&$4.32$&&$4.77$&$3.81$&$0.97$&&$4.77$&$4.19$&$0.38$\\
\ItalianC&$4.95$&$0.98$&$3.97$&&$4.73$&$3.60$&$1.13$&&$4.73$&$4.13$&$0.52$\\
\bottomrule
\end{tabular}}
\caption{Channel code entropies and absolute redundancy.  
By removing predictions, we could reduce the absolute redundancy of the PolyMorph channel code, however, this choice would reduce 
the speller efficiency. 
}\label{table:rate_and_redundancy} 
\end{center}
\end{table}

PolyMorph exhibits an increased OCM with respect to P3Speller whether the spelt sentences are contained into the KB ($\mu \approx 6.38$, $\sigma \approx0.81$ greater than  
$1.43$, Wilcoxon test: $p$-value $=0.015626$) or not 
 ($\mu\approx2.86$, $\sigma\approx0.35$ greater than  $1.43$, Wilcoxon test: $p$-value $=0.015626$). 
With no doubt, the former case is the most favorable one as it causes an average improvement of about $350\%$ with respect to P3Speller, but the latter case has a still remarkable mean enhancement of about $100\%$. 
The amplitude of the improvement depends, as expected, on the user language with peaks of about $430\%$ for Finnish $\Ain$ (former case) and $150\%$ for spelling Hungarian $\Aout$ (latter case) and  
there is a high correlation between the measured OCM and the average characters per word ($\rho=0.76$ 
and, if we do not consider \ItalianC, $\rho=0.98$). 
Intriguingly, there is no apparent relation between the OCM gain in spelling either $\Ain$'s  or $\Aout$'s  (i.e., sentences in the KB and sentences outside the KB) ($\rho\approx 0.04$ and, if we do not consider \ItalianC, $\rho\approx-0.23$). 
As it concerns the former condition test set, the worst case language appears to be the one of the 
Italian curated phrasebook. This is due to the sentence lengths in the phrasebooks: 
since the sentences in the Italian curated phrasebook are shorter on average than those in the other 
phrasebooks (see Table~\ref{table:statistics}), the SSP based 
prediction decreases its effectiveness. 

\begin{figure}[!h]
\begin{center}
\resizebox{\columnwidth}{!}{%
\begin{tikzpicture}[scale=0.80]
\begin{axis}
[
width=18cm,
height=6.5cm,
ymajorgrids,
ybar,
xlabel = Languages,
nodes near coords, 
every node near coord/.append style={rotate=90, anchor=west},
symbolic x coords={English,German,French,Italian,Finnish,Hungarian,{Italian Curated}},
xtick={English,German,French,Italian,Finnish,Hungarian,{Italian Curated}},
xticklabels={\English,\German,\French,\Italian,\Finnish,\Hungarian,\ItalianC},
ylabel = {Output Characters per Minute},
ymin = 0,
ymax=9.3,
legend columns=-1,
legend style={/tikz/every even column/.append style={column sep=0.3cm},at={(0.5,-0.3)}, anchor=north},
point meta=rawy,
]
\addplot[P3SpellerColor,fill=P3SpellerColor] table[y=CPM_P3S] from \insilicodata;
\addlegendentry{P3Speller};
\addplot[nKBColor,fill=nKBColor] table[y=CPM_nKB] from \insilicodata;
\addlegendentry{PolyMorph - $\Aout$};
\addplot[KBColor,fill=KBColor] table[y=CPM_KB] from \insilicodata;
\addlegendentry{PolyMorph - $\Ain$};
\end{axis}
\end{tikzpicture}}
\caption{Mean output characters per minute for \emph{in-silico} experiments when predictions are enabled.}\label{fig:insilico:spelling_time1}
\end{center}
\end{figure}
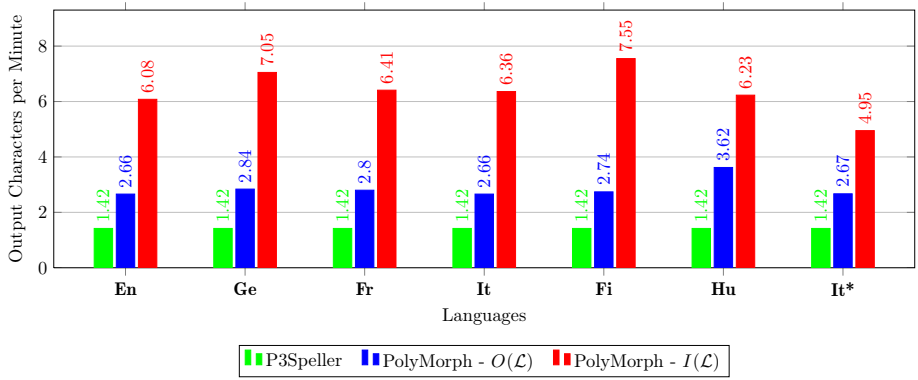

The data measured when predictions were disabled highlighted that 
most of the PolyMorph speed-up is due to the predictions themselves (see Table~\ref{table:charselmin}). However,  
label selections and matrix polymorphism alone significantly
increased OCM with 
respect to P3Speller in spelling both $\Aout$ ($\mu\approx1.88$, $\sigma\approx0.09$ greater than $1.43$, Wilcoxon test: 
$p$-value $=0.015626$)  and 
$\Ain$ ($\mu\approx1.88$, $\sigma\approx0.10$ greater than $1.43$, Wilcoxon test: $p$-value $=0.015626$).  

While the use of PolyMorph in place of P3Speller increases OCM,  SM drop from $1.43$ to $\mu\approx1.29$ ($\sigma\approx0.03$), in spelling sentences that are not contained in 
the KB (Wilcoxon test: $p$-value $=0.015626$),  
and to $\mu\approx1.20$ ($\sigma\approx0.04$), in spelling sentences that are in the KB (Wilcoxon test: 
$p$-value $=0.015626$). In both the cases, SM and ISR have a  
high correlation: $-0.995336$ and $-0.998081$ in spelling $\Aout$'s and $\Ain$'s, respectively.

\section{Conclusions and Future Works}


This paper presents a P300-based speller, named PolyMorph, 
which, among the other features, aims at minimizing the size of the selection matrix and predicts the next word to be spelt taking into account 
what have been already selected. 
While word prediction has been already proposed in the literature (e.g., see~\cite{Ryan:2011nx,dalbis2012,Kaufmann:2012vf}), as far as we known,  
sentence-based predictions, 
polymorphic speller matrix, and 
label selection are exclusive features in the field so far. 
In order to implement the last two features, PolyMorph assumes to have a complete knowledge of the user's dictionary. 
We do not consider this constraint particularly restrictive, however, we plan to remove it in the future. 

PolyMorph differs from the other spellers also because it splits each working cycle into 
a suggestion phase and an identification phase. If, from one hand,  
this choice increases the time due to perform a selection, from the other hand, it allows us to have symbols of similar size 
in the selection matrix. 

We carried out some \emph{in-vivo} and \emph{in-silico} experiments. Although these tests 
are limited in number and do not guarantee the same results in 
subjects with disabilities, they furnish a cheering picture and push us to further investigate PolyMorph. 
In particular, they prove, as already noticed in~\cite{Ryan:2011nx} and \cite{Kaufmann:2012vf}, 
that predictive text entries can increase spelling efficiency.  
They also highlight that sentence-based predictions play a crucial role in the OCM improvements made by PolyMorph (see Section~\ref{sec:results}). 
While the OCM measured by spelling for the first time a sentence that is not present in the KB (target sentence B, turn 1) 
is not really impressive with respect to the ones obtained in~\cite{Ryan:2011nx} and \cite{Kaufmann:2012vf} ($\mu\approx2.70$, $\sigma\approx0.67$ vs 
$\mu\approx5.28$, $\sigma\approx1.67$ and $\mu\approx3.83$, $\sigma\approx0.88$, respectively\footnote{The average OCM's are not reported in~\cite{Kaufmann:2012vf}. We computed them as ratio between 
the number of characters in the target sentence (i.e., $45$) and the overall time needed to spell the sentence (Figure 3 in~\cite{Kaufmann:2012vf}).}), 
targeting a sentence that is already contained in the KB (target sentence A, turn 1) or, better, 
a sentence that is ``frequently'' spelt by the subject (both target sentences A and B, turn 2) produces a massive breakthrough in OCM 
($\mu\approx4.02$, $\sigma\approx1.03$, $\mu\approx5.99$, $\sigma\approx1.78$, and $\mu\approx5.76$, $\sigma\approx1.52$, respectively).  
Obviously, a proper comparison of these results should take into account many factors, such as the average size of the spelt words or the length 
of the spelt sentence. However, these data appear to be more significant at the light of the SM's that, in average, 
were $\mu\approx3.71$, $\sigma\approx0.75$ and 
 $\mu\approx2.12$, $\sigma\approx0.52$ 
in the works of Ryan \emph{et al.}~\cite{Ryan:2011nx} and Kaufmann \emph{et al.}~\cite{Kaufmann:2012vf}, 
respectively\footnote{The average SM's are not reported in~\cite{Kaufmann:2012vf}. Since the authors stated that there 
is no significant differences 
between predictive and non-predictive speller about SM, we approximate them as the SM for non-predictive speller, i.e., 
the ratio between 
the number of characters in the target sentence (i.e., $45$) and the overall time needed to spell the sentence 
(Figure 3 in~\cite{Kaufmann:2012vf}).}, and $\mu\approx1.35$, $\sigma\approx0.35$ for all the PolyMorph sessions i.e., 
turns 1 and 2 
for both target sentences A and B. This means that, when the sentence to be spelt is contained in the KB, 
the OCM's obtained by using PolyMorph are comparable to those measured in~\cite{Ryan:2011nx} and \cite{Kaufmann:2012vf} 
(target sentence B, turn 1) and, in some cases, higher (target sentences A and B, turn 2) even if the proposed speller maintains SM's to a fraction of those of its competitors 
(about the $35\%$ of those indicated by Ryan \emph{et al.}~\cite{Ryan:2011nx}  and about $62\%$ of those estimated from the work of Kaufmann \emph{et al.}~\cite{Kaufmann:2012vf}). 
 
Another major result obtained by PolyMorph concerns user accuracy.  
%
%
In contrast to the spellers presented in~\cite{Ryan:2011nx} and \cite{Kaufmann:2012vf} which either reduce it (the former) or preserve it (the latter)  
with respect to P3Speller,  PolyMorph increased AC from $272/311$ 
on P3Speller (the average probability of correct selection is $\mu\approx0.89$, $\sigma\approx0.09$)  to $75/79$ ($\mu\approx0.96$, $\sigma\approx0.06$), 
$52/52$ ($\mu=1.00$, $\sigma=0.00$), $114/118$ ($\mu\approx0.97$, $\sigma\approx0.06$), and $53/55$ ($\mu\approx0.97$, $\sigma\approx0.06$)  
in spelling target sentence A, turn 1 and 2, and target sentence B, turn 1 and 2, respectively.
This trend was confirmed also by EC which went from $41/230$ (the probability of wrong selection per character is $\mu\approx0.18$, $\sigma\approx 0.21$) 
to $4/230$ ($\mu\approx0.02$, $\sigma\approx0.03$), $0/230$ ($\mu\approx0.00$, $\sigma\approx0.00$), $4/230$ ($\mu\approx0.02$, $\sigma\approx0.04$), and 
$2/230$ ($\mu\approx0.01$, $\sigma\approx0.02$), respectively. 



The reported results  are exciting, but they strictly depend on the sentence to be spelt and, in particular, on its relative frequency in the subject language. Since 
the subject language itself is user dependent and it has many chances to be affected by the spelling 
device (e.g., see the case of SMS~\cite{ahmed2010,Ong2011,Dansieh2011}), 
we plan to perform long-term experiments to evaluate the impact of the proposed techniques on a real spelling environment. 

We observed that almost no error had been accounted during PolyMorph sessions. Thus,   
in the future, we will investigate the relation between time parameters, output characters per minute, and accuracy. 
We will also remove some of the 
constraints of the current version, for instance, by allowing users to dynamically enrich the 
vocabulary. Finally, we would like to integrate in PolyMorph a knowledge graph 
that provides the most likely sentences according to the situation. 
Such a context-aware mechanism will increase the odd of predict the word that is going to be spelt 
by the user and, as a consequence, will have positive effects on the OCM.

\section*{Acknowledgements}
This work has been partially supported by Istituto Nazionale di Alta Matematica
(INdAM) and by University of Trieste as one of the outcomes of the project FRA 2014 ``Learning specifications and robustness in signal analysis (with a case study related to health care)''. 

\bibliographystyle{plain}
\bibliography{biblio}

\appendix
\section{Experimental data}


\begin{table}[!h]
\begin{center}
\resizebox{\columnwidth}{!}{%
\subfloat[Ouput characters per minute (OCM)\label{table:invivo_OCM}]{
\begin{tabular}{lccccc}
\toprule
\headcol & \multicolumn{3}{c}{Sentence A}& \multicolumn{2}{c}{Sentence B}\\
 \rulefiller\cmidrule(r){2-4} \cmidrule(r){5-6}
\headcol Subj &  PM $1$& PM $2$ & P3S & PM $1$ &  PM $2$ \\
\midrule
\subject{1}&$6.48$&$8.94$&$2.51$&$4.16$&$8.94$\\
\rowcol \subject{2}&$2.85$&$5.03$&$1.07$&$2.35$&$5.03$\\
\subject{3}&$4.11$&$5.65$&$1.06$&$2.64$&$5.65$\\
\rowcol \subject{4}&$2.76$&$3.79$&$0.78$&$1.78$&$3.79$\\
\subject{5}&$3.87$&$5.32$&$0.88$&$2.49$&$3.80$\\
\rowcol \subject{6}&$3.76$&$8.94$&$1.07$&$3.06$&$6.39$\\
\subject{7}&$3.94$&$6.92$&$1.45$&$3.23$&$6.92$\\
\rowcol \subject{8}&$4.37$&$5.01$&$1.54$&$2.81$&$6.02$\\
\subject{9}&$3.66$&$4.28$&$0.93$&$2.35$&$5.03$\\
\rowcol \subject{10}&$4.37$&$6.02$&$1.15$&$2.12$&$6.02$\\
\bottomrule
\end{tabular}}
\hskip5mm
\subfloat[Selections per minute (SM)\label{table:invivo_SM}]{
\begin{tabular}{lccccc}
\toprule
\headcol & \multicolumn{3}{c}{Sentence A}& \multicolumn{2}{c}{Sentence B}\\
 \rulefiller\cmidrule(r){2-4} \cmidrule(r){5-6}
\headcol Subj &  PM $1$& PM $2$ & P3S & PM $1$ &  PM $2$ \\
\midrule
\subject{1}&$1.97$&$1.94$&$2.51$&$1.99$&$1.94$\\
\rowcol \subject{2}&$1.12$&$1.09$&$1.25$&$1.13$&$1.09$\\
\subject{3}&$1.25$&$1.23$&$1.43$&$1.26$&$1.23$\\
\rowcol \subject{4}&$0.84$&$0.82$&$0.91$&$0.85$&$0.82$\\
\subject{5}&$1.18$&$1.16$&$1.34$&$1.19$&$1.16$\\
\rowcol \subject{6}&$1.96$&$1.94$&$2.51$&$2.00$&$1.94$\\
\subject{7}&$1.54$&$1.50$&$1.83$&$1.54$&$1.50$\\
\rowcol \subject{8}&$1.33$&$1.31$&$1.54$&$1.34$&$1.31$\\
\subject{9}&$1.11$&$1.12$&$1.25$&$1.13$&$1.09$\\
\rowcol \subject{10}&$1.33$&$1.31$&$1.54$&$1.39$&$1.31$\\
\bottomrule
\end{tabular}}}
\caption{
Output characters per minute (OCM) and selections per minute (SM) for PolyMorph turn 1 (PM $1$), PolyMorph turn 2 (PM $2$),  
and P3Speller (P3S)  for \emph{in-vivo} experiments.\label{table:OCMS}}
\end{center}
\end{table}

\begin{table}[!h]
\begin{center}
\begin{tabular}{lccccc}
\toprule
\headcol & \multicolumn{3}{c}{Sentence A}& \multicolumn{2}{c}{Sentence B}\\
 \rulefiller\cmidrule(r){2-4} \cmidrule(r){5-6}
\headcol Subj &  PM $1$& PM $2$ & P3S & PM $1$ &  PM $2$ \\
\midrule
\subject{1}&$11.71$&$12.00$&$12.00$&$11.55$&$12.00$\\
\rowcol \subject{2}&$11.67$&$12.00$&$12.00$&$11.55$&$12.00$\\
\subject{3}&$11.71$&$12.00$&$12.00$&$11.55$&$12.00$\\
\rowcol \subject{4}&$11.71$&$12.00$&$12.00$&$11.55$&$12.00$\\
\subject{5}&$11.71$&$12.00$&$12.00$&$11.55$&$12.00$\\
\rowcol \subject{6}&$11.83$&$12.00$&$12.00$&$11.47$&$12.00$\\
\subject{7}&$11.56$&$12.00$&$12.00$&$11.55$&$12.00$\\
\rowcol \subject{8}&$11.71$&$12.00$&$12.00$&$11.55$&$12.00$\\
\subject{9}&$11.71$&$11.67$&$12.00$&$11.55$&$12.00$\\
\rowcol \subject{10}&$11.71$&$12.00$&$12.00$&$11.07$&$12.00$\\
\bottomrule
\end{tabular}
\caption{Intensifications per selection and repetition (ISR) for PolyMorph turn 1 (PM $1$),  turn 2 (PM $2$),  
and P3Speller (P3S)  for \emph{in-vivo} experiments.\label{table:invivo_ISR}}
\end{center}
\end{table}

\pgfplotstableread{
Lang	P3S	PM_Pred_Out	PM_Pred_In	PM_no_Pred_Out	PM_no_Pred_In
English	12.00	11.70	12.81	10.58	10.55
German	12.00	11.34	12.77	10.41	10.40
French	12.00	11.47	12.85	10.35	10.38
Italian	12.00	11.40	12.77	10.30	10.31
Finnish	12.00	10.81	12.48	9.78	9.78
Hungarian	12.00	11.18	11.80	10.40	10.35
{Italian Curated}	12.00	10.66	11.44	9.92	9.92

}\insiliconis

\begin{figure}[!h]
\begin{center}
\resizebox{\columnwidth}{!}{%
\begin{tikzpicture}[scale=0.80]
\begin{axis}
[
width=18cm,
height=6cm,
ymajorgrids,
ybar,
ymode = log,
xlabel = Languages,
nodes near coords, 
every node near coord/.append style={rotate=90, anchor=west},
symbolic x coords={English,German,French,Italian,Finnish,Hungarian,{Italian Curated}},
xtick={English,German,French,Italian,Finnish,Hungarian,{Italian Curated}},
xticklabels={\English,\German,\French,\Italian,\Finnish,\Hungarian,\ItalianC},
ylabel style={align=center}, 
ylabel=Intensifications per\\Selection and Repetition,
ymin = 0,
ymax=13.8,
legend columns=-1,
legend style={/tikz/every even column/.append style={column sep=0.3cm},at={(0.5,-0.3)}, anchor=north},
point meta=rawy,
]
\addplot[P3SpellerColor,fill=P3SpellerColor] table[y=P3S] from \insiliconis;
\addlegendentry{P3Speller};
\addplot[nKBColor,fill=nKBColor] table[y=PM_Pred_Out] from \insiliconis;
\addlegendentry{PolyMorph - $\Aout$};
\addplot[KBColor,fill=KBColor] table[y=PM_Pred_In] from \insiliconis;
\addlegendentry{PolyMorph - $\Ain$};
\end{axis}
\end{tikzpicture}}
\caption{Intensifications per selection and repetition for \emph{in-silico} experiments when predictions are enabled.}\label{fig:insilico:ISR_Pred}
\end{center}
\end{figure}
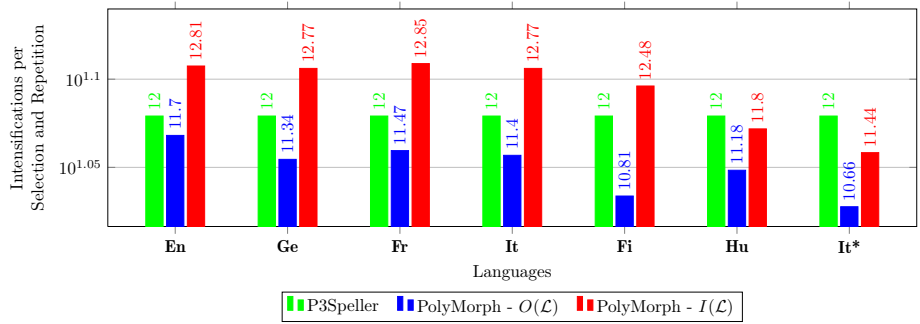

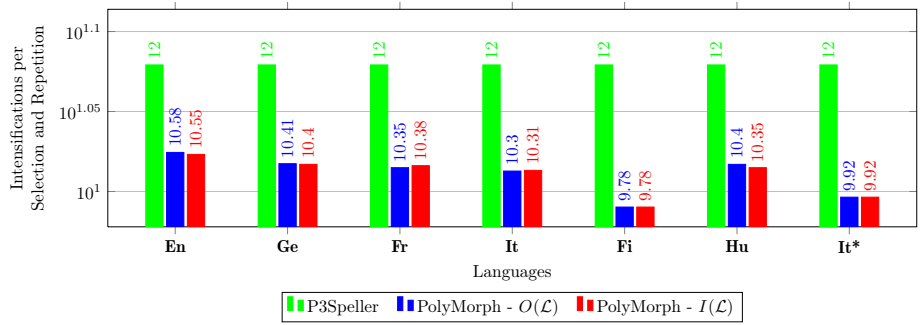
\begin{figure}[!h]
\begin{center}
\resizebox{\columnwidth}{!}{%
\begin{tikzpicture}[scale=0.80]
\begin{axis}
[
width=18cm,
height=6cm,
ymajorgrids,
ybar,
ymode = log,
xlabel = Languages,
nodes near coords, 
every node near coord/.append style={rotate=90, anchor=west},
symbolic x coords={English,German,French,Italian,Finnish,Hungarian,{Italian Curated}},
xtick={English,German,French,Italian,Finnish,Hungarian,{Italian Curated}},
xticklabels={\English,\German,\French,\Italian,\Finnish,\Hungarian,\ItalianC},
ylabel style={align=center}, 
ylabel=Intensifications per\\Selection and Repetition,
ymin = 0,
ymax=13,
legend columns=-1,
legend style={/tikz/every even column/.append style={column sep=0.3cm},at={(0.5,-0.3)}, anchor=north},
point meta=rawy,
]
\addplot[P3SpellerColor,fill=P3SpellerColor] table[y=P3S] from \insiliconis;
\addlegendentry{P3Speller};
\addplot[nKBColor,fill=nKBColor] table[y=PM_no_Pred_Out] from \insiliconis;
\addlegendentry{PolyMorph - $\Aout$};
\addplot[KBColor,fill=KBColor] table[y=PM_no_Pred_In] from \insiliconis;
\addlegendentry{PolyMorph - $\Ain$};
\end{axis}
\end{tikzpicture}}
\caption{Intensifications per selection and repetition for \emph{in-silico} experiments when predictions are disabled.}\label{fig:insilico:ISR_no_Pred}
\end{center}
\end{figure}

\begin{table}[!h]
\begin{center}
\resizebox{\columnwidth}{!}{%
\begin{tabular}{lcccccccccc}
\toprule
\headcol & \multicolumn{2}{c}{P3S}& \multicolumn{3}{c}{PM - Pred.} & \multicolumn{3}{c}{PM - No Pred.}\\
 \rulefiller\cmidrule(r){2-3} \cmidrule(r){4-6}\cmidrule(r){7-9}
\headcol $\mathcal{L}$&Any&$|\nrows*\ncols|$&$\Aout$&$\Ain$&$\max{|\nrows*\ncols|}$&$\Aout$&$\Ain$&$\max{|\nrows*\ncols|}$\\
\midrule
\English&$12.00$&$6\times 6$&$11.70$&$12.81$&$7\times 6$&$10.58$&$10.55$&$6\times 6$\\
\rowcol \German&$12.00$&$6\times 6$&$11.34$&$12.77$&$7\times 6$&$10.41$&$10.40$&$6\times 6$\\
\French&$12.00$&$6\times 6$&$11.47$&$12.85$&$7\times 6$&$10.35$&$10.38$&$6\times 6$\\
\rowcol \Italian&$12.00$&$6\times 6$&$11.40$&$12.77$&$7\times 6$&$10.30$&$10.31$&$6\times 6$\\
\Finnish&$12.00$&$6\times 6$&$10.81$&$12.48$&$7\times 6$&$9.78$&$9.78$&$6\times 6$\\
\rowcol \Hungarian&$12.00$&$6\times 6$&$11.18$&$11.80$&$7\times 6$&$10.40$&$10.35$&$6\times 6$\\
\ItalianC&$12.00$&$6\times 6$&$10.66$&$11.44$&$6\times 6$&$9.92$&$9.92$&$6\times 5$\\
\bottomrule
\end{tabular}}
\caption{Intensifications per Selection and Repetition (ISR)  for \emph{in-silico} experiments. 
The presence of the prediction symbols in the selection matrix may increase the number of rows $\nrows$. 
However, whenever the sentence to be spelt is not contained in the KB
the ISR of PolyMorph is smaller than that of P3Speller. 
The data observed when we disabled predictions 
(i.e. PM - No Pred.) highlight the advantage of using PolyMorph over P3Speller in term of intensifications required to 
spell a sentence.}\label{table:ISR} 
\end{center}
\end{table}

\begin{figure}[!h]
\begin{center}
\resizebox{\columnwidth}{!}{%
\begin{tikzpicture}[scale=0.80]
\begin{axis}
[
width=18cm,
height=6cm,
ymajorgrids,
ybar,
xlabel = Languages,
nodes near coords, 
every node near coord/.append style={rotate=90, anchor=west},
symbolic x coords={English,German,French,Italian,Finnish,Hungarian,{Italian Curated}},
xtick={English,German,French,Italian,Finnish,Hungarian,{Italian Curated}},
xticklabels={\English,\German,\French,\Italian,\Finnish,\Hungarian,\ItalianC},
ylabel = {Output Characters per Minute},
ymin = 0,
ymax=3,
legend columns=-1,
legend style={/tikz/every even column/.append style={column sep=0.3cm},at={(0.5,-0.3)}, anchor=north},
point meta=rawy,
]
\addplot[P3SpellerColor,fill=P3SpellerColor] table[y=CPM_P3S] from \insilicodatanohints;
\addlegendentry{P3Speller};
\addplot[nKBColor,fill=nKBColor] table[y=CPM_nKB] from \insilicodatanohints;
\addlegendentry{PolyMorph - $\Aout$};
\addplot[KBColor,fill=KBColor] table[y=CPM_KB] from \insilicodatanohints;
\addlegendentry{PolyMorph - $\Ain$};
\end{axis}
\end{tikzpicture}}
\caption{Mean output characters per minute for \emph{in-silico} experiments when predictions are disabled.}\label{fig:insilico:spelling_time1_no_hints}
\end{center}
\end{figure}
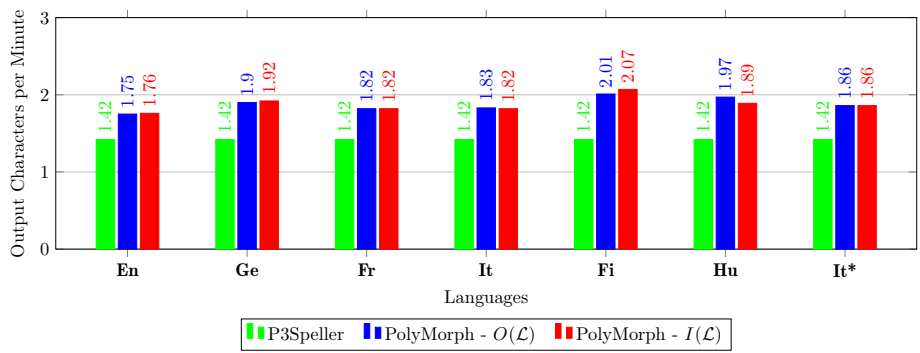

\begin{table}[!h]
\begin{center}
\resizebox{\columnwidth}{!}{%
\subfloat[Mean output characters per minute\label{table:charmin}]{
\begin{tabular}{lccccccccc}
\toprule
\headcol & \multicolumn{1}{c}{P3S}& \multicolumn{2}{c}{PM - Pred.} & \multicolumn{2}{c}{PM - No Pred.}\\
 \rulefiller\cmidrule(r){2-2} \cmidrule(r){3-4}\cmidrule(r){5-6}
\headcol $\mathcal{L}$&Any&$\Aout$&$\Ain$&$\Aout$&$\Ain$\\
\midrule
\English&$1.43$&$2.66$&$6.08$&$1.75$&$1.76$\\
\rowcol \German&$1.43$&$2.84$&$7.05$&$1.9$&$1.92$\\
\French&$1.43$&$2.8$&$6.41$&$1.82$&$1.82$\\
\rowcol \Italian&$1.43$&$2.66$&$6.36$&$1.83$&$1.82$\\
\Finnish&$1.43$&$2.74$&$7.55$&$2.01$&$2.07$\\
\rowcol \Hungarian&$1.43$&$3.62$&$6.23$&$1.97$&$1.89$\\
\ItalianC&$1.43$&$2.67$&$4.95$&$1.86$&$1.86$\\
\bottomrule
\end{tabular}}\hskip5mm
\subfloat[Mean selections per minute\label{table:selmin}]{
\begin{tabular}{lccccccccc}
\toprule
\headcol & \multicolumn{1}{c}{P3S}& \multicolumn{2}{c}{PM - Pred.} & \multicolumn{2}{c}{PM - No Pred.}\\
 \rulefiller\cmidrule(r){2-2} \cmidrule(r){3-4}\cmidrule(r){5-6}
\headcol $\mathcal{L}$&Any&$\Aout$&$\Ain$&$\Aout$&$\Ain$\\
\midrule
\English&$1.43$&$1.25$&$1.17$&$1.60$&$1.60$\\
\rowcol\German&$1.43$&$1.28$&$1.17$&$1.62$&$1.62$\\
\French&$1.43$&$1.27$&$1.17$&$1.62$&$1.62$\\
\rowcol\Italian&$1.43$&$1.27$&$1.17$&$1.63$&$1.63$\\
\Finnish&$1.43$&$1.32$&$1.19$&$1.70$&$1.70$\\
\rowcol\Hungarian&$1.43$&$1.29$&$1.24$&$1.62$&$1.63$\\
\ItalianC&$1.43$&$1.34$&$1.27$&$1.68$&$1.68$\\
\bottomrule
\end{tabular}}}
\caption{Mean output characters per minute (OCM) and selections per minute (SM) for \emph{in-silico} experiments.}\label{table:charselmin}
\end{center}
\end{table}

\end{document}